# Magnetic instabilities in quasi-one-dimensional Cr-based material

A. Galluzzi[1,2], G. Cuono[3], A. Romano[1,2], J. Luo[4,5,6], C. Autieri[3,2], C. Noce[1,2] and M. Polichetti[1,2]

[1]Department of Physics "E.R. Caianiello", University of Salerno, via Giovanni Paolo II, 132, Fisciano (SALERNO), I-84084, Italy

[2]CNR-SPIN Salerno, via Giovanni Paolo II, 132, Fisciano (SALERNO), I-84084, Italy

[3]International Research Centre Magtop, Institute of Physics, Polish Academy of Sciences, Aleja Lotników 32/46, PL-02668 Warsaw, Poland

[4]Beijing National Laboratory for Condensed Matter Physics and Institute of Physics, Chinese Academy of Sciences, Beijing 100190, China

[5]Songshan Lake Materials Laboratory, Dongguan, Guangdong 523808, China

[6]School of Physical Sciences, University of Chinese Academy of Sciences, Beijing 100190, China

**Abstract**

The magnetic response of a $K_2Cr_3As_3$ sample has been studied by means of dc magnetization measurements as a function of magnetic field (H) at different temperatures ranging from 5 K up to 300 K. Looking at the magnetic hysteresis loops m(H), a diamagnetic behavior of the sample has been inferred at temperatures higher than 60 K, whereas at lower temperatures the sample shows a hysteresis loop compatible with the presence of ferrimagnetism. Moreover, several spike-like magnetization jumps, both positive and negative, have been observed at certain fields in the range -1000 Oe < H < 1000 Oe, regardless of the temperature considered. The field position of the magnetization jumps has been studied at different temperatures, and their distribution can be described by a Lorentzian curve.

## 1. Introduction

In the last few years, the quasi-one-dimensional chromium-based superconductor $K_2Cr_3As_3$[1] has been extensively studied. Indeed, it exhibits many interesting properties, regarding the interplay between magnetism, structural properties, superconductivity and quasi-one-dimensionality, shared with other compounds of the family $A_2Cr_3As_3$ (where A=Na[2], Rb[3], Cs[4] other than K) and with related compounds belonging to the series $ACr_3As_3$ [5,6]. For completeness, we mention that compounds having the same crystal structure but based on molybdenum $A_2Mo_3As_3$ [7–10] have been discovered and other Cr-based superconductors, such as the CrAs, have been recently intensively studied[11–16]. Regarding the magnetic properties of $K_2Cr_3As_3$ and similar materials, both theoretical[17,18] and experimental[1] studies suggest that the ground state is non-magnetic. Theoretically, the nonmagnetic ground state of $K_2Cr_3As_3$ was attributed to the triangular configuration of the chains composed of chromium atoms which leads to frustration. Experimentally, it has been found that these systems are paramagnetic with small hysteresis loop at low temperature attributed to defects or magnetic impurities[4], probably due to the presence of $KCr_3As_3$, which instead is magnetic and presents a cluster spin-glass state[5]. Nevertheless, there are theoretical speculations that show that $K_2Cr_3As_3$ possesses strong magnetic fluctuations and is close to a non-collinear magnetic ground state in-out[19]. Besides, nuclear quadrupole resonance indicates that moving along the series A=Na, $Na_{0.75}K_{0.25}$, K, Rb, the system tends to approach a possible ferromagnetic quantum critical point[20]. However, neutron scattering measurements have established the presence of phonon instabilities related to structural distortions in $K_2Cr_3As_3$[21]. These distortions from the triangular structure make the system no longer frustrated. Very recently, we have shown that there is a strong interplay between the structural distortions and the magnetism, and we have found, for the distorted structure, a collinear intrachain ferrimagnetic ground state for the $KCr_3As_3$ and an instability towards the same collinear intrachain ferrimagnetic phase for the $K_2Cr_3As_3$[22,23]. Assuming an intrachain ferrimagnetism, the system could exhibit a long-range ferrimagnetism or spin-glass phase depending on the interchain magnetic coupling. Our results predict that $K_2Cr_3As_3$ is nonmagnetic but close to a long-range ferrimagnetic phase, while the $KCr_3As_3$ goes in a spin-glass phase. Interestingly, when the $K_2Cr_3As_3$ is considered, the application of a strain can drive the compound from one phase to the other, since the non-magnetic phase and the ferrimagnetic one are close in energy[23].

Here, we study a Cr-based quasi-one-dimensional sample containing $K_2Cr_3As_3$ by means of dc magnetization measurements as a function of magnetic field H at different temperatures, ranging from 5 K up to 300 K. Looking at the magnetic hysteresis, we find that at low temperatures the sample shows a ferrimagnetic behavior while at temperatures higher than 60 K a diamagnetic behavior appears. Several spike-like magnetization jumps have been found in a range -1000 Oe < H < 1000 Oe and this can be explained as due to the magnetic instabilities. The paper is organized as follows: in the next section the experimental results are presented, section 3 is devoted to the results and discussion whereas the last one contains the conclusions.

## 2. Experimental details

A needle-shaped $K_2Cr_3As_3$ single crystal having length and thickness equal to 2.5 mm and 0.1 mm, respectively, has been analyzed. The fabrication details are reported elsewhere[1,24]. The composition of our sample is $K_{1.81}Cr_{3.57}As_3$. Therefore, the ratio between the K and the Cr is 0.507 which is almost halfway between 0.333 (ratio of the $KCr_3As_3$) and 0.666 (ratio of the $K_2Cr_3As_3$). So, we can assume that the sample is composed by both sizeable part of $KCr_3As_3$ and $K_2Cr_3As_3$ as already reported in Ref.[25].

The sample has been characterized in a dc magnetic field applied perpendicular to its length. In particular, the dc magnetic moment as a function of the field m(H) has been measured using a Quantum Design PPMS-9T equipped with a VSM option. To avoid the effect on the sample response due to the residual trapped field inside the PPMS dc magnet[26], this field was reduced below $1 \times 10^{-4}$ T[27]. For what concerns the M(H) measurements, the sample was first cooled down to the measurement temperature in zero field and thermally stabilized for at least 20 min[28,29]. Then, the field was ramped up to +9 T, then back to −9 T, and finally to +9 T again to acquire the complete hysteresis loop[30,31].

## 3. Results and discussion

In order to study the magnetic response of the sample, the magnetic moment m has been measured as a function of the magnetic field H in the temperature range from 5 K up to 300 K. In the main panel of Figure 1, some of the measured m(H) curves have been reported. It can be noted that up to 55 K the m(H) curves show a magnetic behavior which could be associated with the paramagnetism, ferromagnetism or ferrimagnetism phenomenology while for T = 60 K the sample is in a diamagnetic state (see inset(a) of Figure 1). This suggests a magnetic transition temperature at $T^* \approx 58$ K. In the inset(b) of Figure 1, the m(H) at T = 30 K has been reported as an example, focusing on the region near zero field for evaluating the possible existence of coercivity. It is well visible how the coercive field is different from zero ($H_c \approx 100$ Oe) so discarding the possibility of a paramagnetic behavior. It is worth to underline that $H_c \neq 0$ Oe is obtained for all the temperatures below 60 K. Since for $T \geq 60$ K the sample is diamagnetic, a coexistence of two sublattices characterized by two opposite magnetic ordering is plausible. This situation is not in contrast with a ferrimagnetic behavior, in agreement with the results reported in Refs.[22,23].

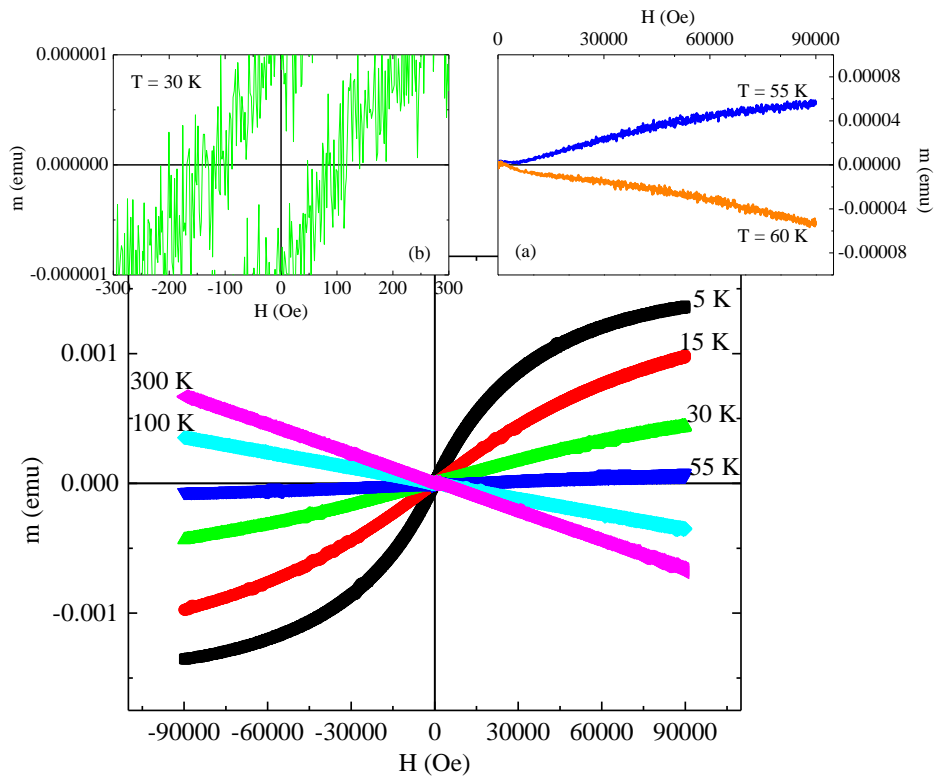

**Figure 1** m(H) curves at different temperatures. Inset(a): m(H) curves at T = 55 K and T = 60 K. Inset(b): the region near zero field has been magnified for the m(H) curve at T = 30 K.

An interesting feature of the m(H) curves reported in Figure 1 is visible by focusing on the magnetic response of the sample in the field range -1000 Oe < H < +1000 Oe. In fact, several spike-like magnetization jumps, both positive and negative, are observed at different fields, regardless of the temperature considered (see Figure 2). The red circles drawn in the panels of Figure 2 individuate the field position of the magnetization jumps. It is important to highlight that the magnetization jumps are visible for all the reported temperatures independently of the magnetic state that characterizes the sample and independently of the positive or negative values of the applied magnetic field.

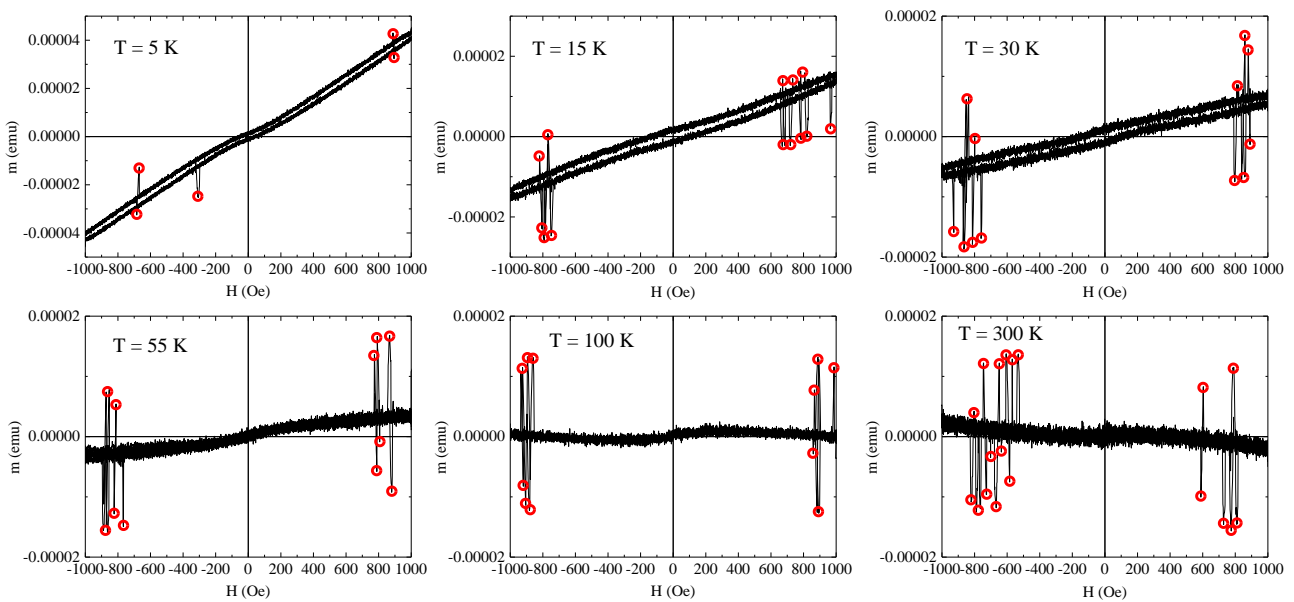

**Figure 2** m(H) curves in the range -1000 Oe < H < +1000 Oe. Several spike-like magnetization jumps (indicated by open red circles) can be observed at different fields for all the reported temperatures.

The field positions of the magnetization jumps ($H_{jump}$) have been extracted from all the panels of Figure 2 and a histogram has been constructed by setting the frequency expressed as a percentage (%) on y axis and the $H_{jump}$ values (taken in absolute value) on x axis. The result is reported in Figure 3. Each bin has a width of 50 Oe with the highest one ranging between 800 Oe and 850 Oe showing the 30% of presence respect to the total values. It is important to note that the interval ranging between 700 Oe and 900 Oe contains more than the 85% of total values indicating a narrow field region where most of the magnetization jumps occur. This could suggest the presence of magnetic processes activating in correspondence to a specific magnetic energy, allowing the magnetization to jump from a magnetic state to another one. The distribution reported in Figure 3 can be well fitted by a Lorentzian curve (peak average ≈ 820 Oe) which usually characterizes the behavior observed in the absorption spectra.

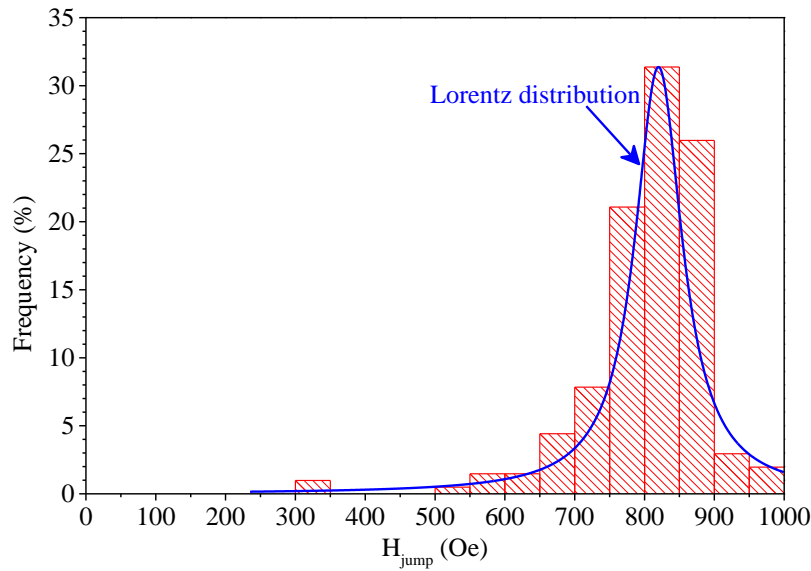

**Figure 3** Distribution of the $H_{jump}$ values. The blue curves represent the fit of the data with a Lorentz distribution curve.

**Conclusions**

To study the magnetic response of a $K_2Cr_3As_3$ sample, dc magnetization measurements have been performed as a function of the magnetic field at different temperatures ranging from 5 K up to 300 K. Analyzing the magnetic hysteresis loops m(H), we have found a magnetic transition at T ≈ 60 K from a diamagnetic state (T ≥ 60 K) to a paramagnetic response (T < 60 K) compatible with an overall ferrimagnetic behavior of the sample. We underline that ours are bulk measurements so the results obtained are not due to surface effects. Moreover, focusing on the magnetic response of the sample in the field range -1000 Oe < H < +1000 Oe, we have found several spike-like magnetization jumps in the entire temperature range. The field positions of the magnetization jumps have been extracted at the different temperatures, reported in a histogram, and then fitted by a Lorentzian distribution curve. We have found that more than 85% of values are included in the field range 700 Oe ≤ H ≤ 900 Oe suggesting that the lattice needs a specific magnetic energy for allowing the magnetization to jump from a magnetic state to another one.

**Acknowledgments**

G.C. and C.A. are supported by the Foundation for Polish Science through the International Research Agendas program co-financed by the European Union within the Smart Growth Operational Programme. G. C. acknowledges financial support from "Fondazione Angelo Della Riccia".